\documentclass[preprint,showpacs,preprintnumbers,amsmath,amssymb]{revtex4}

\usepackage{graphicx}% Include figure files
\usepackage{dcolumn}% Align table columns on decimal point
\usepackage{bm}% bold math

\begin{document}

\title{The equation of state for the nuclear matter \\and the properties of the neutron star}

\author{Chang-Geng Liu and Bao-Xi Sun}

\affiliation{Institute of Theoretical Physics, College of Applied
Sciences, Beijing University of Technology, Beijing 100022, China }

\begin{abstract}
The equation of state for the $\beta$ stable nuclear matter is
calculated numerically, and then the Tolman-Oppenheimer-Volkov(TOV)
equation for the structure of the neutron star is solved in the
fourth-order Runge-Kutta algorithm. It shows the mass and radius of
the neutron star are functions of the central density of the neutron
star and a maximum mass of 1.932 solar masses with a corresponding
radius of 9.340km is obtained. Considering the equation of state of
the nuclear matter must obey the causality, a new factor $c$ is
added in the nuclear potential energy formula. Therefore, with a new
equation of state for the $\beta$ stable nuclear matter when
$c=0.15$, a new maximum mass of 1.440 solar masses with a radius of
9.765km for the neutron star is obtained. Finally, the contribution
of the cosmological constant to the structure of the neutron star is
discussed, and we find the cosmological constant has minimal or
negligible influence on the properties of the neutron star.\\ Key
Words:\ \ Nuclear matter, Equations of state, Neutron stars\\
\end{abstract}
\maketitle

{\it  Chang-Geng Liu is a diligent and excellent student. This
article is based on his graduate thesis, which is chosen as one of
the excellent theses in Beijing University of Technology in 2007.---
Bao-Xi Sun}

\section{\normalsize introduction}
Since the first pulsar was discovered in 1967, it has been generally
believed that the pulsar is a rapidly rotating neutron star endowed
with a strong magnetic field. Thereafter, studies upon the
properties of neutron stars gained wide concerns.

Until now, many different views on the structure of the neutron
star, especially the composition of  the matter in the core of the
neutron star, remain among theoretical physicists because it is
difficult to analyze the components of the neutron star matter from
the present astronomical observations\cite{He00}. The core of the
neutron star, where the density is very large and thence the kaon
condensation could happen, possibly consists of the strange hadronic
matter including hyperons and there could also be the large-size
quark matter in it. In 2006, it was concluded by Ozel that the
condensates and unconfined quarks do not exist in the cores of
neutron stars, because the mass and radius of the neutron star EXO
0748-676 rule out all the soft equations of state of the neutron
star matter\cite{Oz06}. However, Alford et al. stated that the quark
matter can be as stiff as the nuclear matter, and the corresponding
hybrid or quark stars can reach a mass of 2.0 solar
masses\cite{Al07}.

In the study of the structure of the neutron star, except the
non-relativistic Brueckner theory\cite{Zu04,Zh04}, the Quantum
Hadrodynamics (QHD) based on the Walecka model has made great
progress in the past 30 years\cite{Se86,Re89,Ri96,Me06}, which is
also used to calculate the equation of state for  the neutron star
matter \cite{Sh98,Ta04}. To soften the equation of state for the
neutron star matter, some research groups add self-interaction terms
of mesons\cite{Mu96} in the Lagrangian density of QHD. At the
meantime, considering the pair-correlation between nucleons, some
scientists have established the relativistic Harteee-Bogliubov
method\cite{Se02,Me02}. In addition, some others have developed the
chiral hadronic model of QHD due to the chiral symmetry spontaneous
breaking\cite{Fu97,Pa99}.

In this article, the nuclear model in Refs.~\cite{Pr96,Si03} is
utilized to calculate the equation of state for the neutron star
matter and study further the properties of neutron stars although a
similar work has been done by V. P. Psonis et al., where the
relation between the asymmetric energy of the nuclear matter and the
properties of the neutron star is discussed within the same
model\cite{Ps07}.
The equilibrium conditions that the chemical potentials of protons,
neutrons and electrons in the neutron star matter will be discussed
in Section ~\ref{beta}. The equation of state for the neutron star
matter and the requirement that the equation of state must meet to
obey the causality will be discussed in Sections ~\ref{eos} and
~\ref{causality}, respectively, and the numerical results on the
properties of neutron stars are analyzed in Section ~\ref{result}.
The probable influence of the cosmological constant on the neutron
star will be studied in Section ~\ref{cos}. The summary is given in
Section ~\ref{sum}. It only needs some knowledge about quantum
mechanics to calculate the equation of state with this model, so
this work is suitable as a subject for the graduation thesis of an
undergraduate physics major. Actually, this article is just based on
Chang-Geng Liu's graduation thesis.

\section{\label{beta}\normalsize $\beta$ equilibrium conditions}
Because a free neutron will undergo a beta weak decay process
$n\rightarrow p+e^{-}+\overline{v}_{e}$, there are not only
neutrons, but also a small fraction of protons and electrons in the
neutron star matter. The neutron star is neutral. Hence, the density
of the proton is equal to that of the electron, which means their
Fermi momenta are the same as each other, $k_{f,p}{=}k_{f,e}$. When
the neutron decay process takes place, the electron capture reaction
$p+e^{-}\rightarrow n+ v_{e}$ is going on. These two weak reactions
will reach an equilibrium. This equilibrium can be expressed in
terms of the chemical potentials for the three particle species
\begin{equation}
\mu_{n}=\mu_{p}+\mu_{e},
\end{equation}
where the chemical potential for a particle species can be expressed
as
\begin{equation}
\mu_{i}=(k_{f,i}^{2}c^{2}+m_{i}^{2}c^{4})^{\frac{1}{2}},\ \  i=n,p,e
\end{equation}
Thence, the Fermi momentum of the proton in the neutron star matter
which reaches the beta equilibrium can be calculated as
\begin{equation}
k_{f,p}\approx\frac{k_{f,n}^{2}+m_{n}^{2}c^{2}-
m_{p}^{2}c^{2}}{2(k_{f,n}^{2}+m_{n}^{2}c^{2})^{\frac{1}{2}}}.
\end{equation}

\section{\label{eos}\normalsize the equation of state for the neutron star matter}
Considering the influence of interactions between nucleons upon the
equation of state, we mainly refer to Prakash's lectures on the
asymmetric nuclear matter model to construct a simple nuclear
potential energy formula\cite{Pr96,Si03}. The saturation density of
the symmetric nuclear matter $n_{0}=0.16fm^{-3}$, and the binding
energy per nucleon $BE=E/A-m_{N}c^{2}=-16MeV$, where\ $m_{N}$\
represents the mass of the nucleon, and $k_{0}$ is the nuclear
compressibility that is applicable between 200 and 400MeV according
to the experimental data and will be set to 360MeV in the discussion
below. For the symmetric nuclear matter, the density of the neutron
equals that of the proton, $n_{n}{=}n_{p}$, with the total nucleon
density $n{=}n_{n}+n_{p}$. The average energy per nucleon for the
symmetric nuclear matter will be modeled as \cite{Pr96,Si03}
\begin{equation}
\label{eq:energypern}
E(n,0)=\frac{\varepsilon(n)}{n}=m_{N}c^{2}+\frac{3}{5}\frac{k_{f}^{2}}{2m_{N}}
+\frac{A}{2}u+\frac{B}{\sigma+1}u^{\sigma},
\end{equation}
where $u=n/{n_0}$ and $\sigma$ is dimensionless and A and
\vspace{1mm} B have the same dimension with the energy. The first
term denotes the rest mass energy. The second term represents the
average kinetic energy per nucleon with the corresponding Fermi
momentum $k_{f}{=}(3\pi^{2}\hbar^{3}\frac{n}{2})^{\frac{1}{3}}$. At
the saturation density, the average kinetic energy per nucleon is
designated as $<E_{f}^{0}>{=}22.1MeV$, and then the second term can
be rewritten as\ $<E_{f}^{0}>u^{\frac{2}{3}}$. The last two terms in
Eq.~(\ref{eq:energypern}) indicate the mean nuclear potential energy
per nucleon. From the properties of the saturation nuclear matter,
three constraint equations can be obtained to determine A, B and
$\sigma$.
\begin{eqnarray}
<E_{f}^{0}>+\frac{A}{2}+\frac{B}{\sigma+1}&=&BE, \\
\frac{2}{3}<E_{f}^{0}>+\frac{A}{2}+\frac{\sigma B}{\sigma+1}&=&0,\\
\frac{10}{9}<E_{f}^{0}>+A+\sigma B&=&\frac{k_{0}}{9},
\end{eqnarray}
which result in
\begin{eqnarray}
\sigma&=&\frac{k_{0}+2<E_{f}^{0}>}{3<E_{f}^{0}>-9BE},  \\
B&=&\frac{\sigma+1}{\sigma-1}[\frac{1}{3}<E_{f}^{0}>-BE],\\
A&=&BE-\frac{3}{5}<E_{f}^{0}>-B.
\end{eqnarray}
If\ $ k_{0}$\ is determined, A,B and $\sigma $ are fixed
accordingly. When $k_{0}=360MeV$, $\sigma{=}1.922$, $A=-126.986MeV$
and $B=-74.098MeV$.

By  the thermodynamic relation
\begin{equation}
p=n^{2}\frac{d(\varepsilon/n)}{dn}=n\frac{d\varepsilon}{dn}-\varepsilon,
\end{equation}
the pressure in the symmetric nuclear matter can be expressed as
\begin{equation}
\label{eq:prenm}
p(n)=n_{0}[\frac{2}{3}<E_{f}^{0}>u^{\frac{5}{3}}+\frac{A}{2}u^{2}+\frac{\sigma
B}{\sigma+1}u^{\sigma+1}].
\end{equation}

As for the asymmetric nuclear matter, the densities of  the proton
and neutron are not the same any longer. Let us represent the proton
and neutron densities in terms of a symmetry factor
$\alpha=\frac{n_{n}-n_{p}}{n}$ with the proton density
$n_{p}=\frac{1-\alpha}{2}n$ and the neutron density
$n_{n}=\frac{1+\alpha}{2}n$.

As the asymmetric part of the mean energy per nucleon is
approximately proportional to the square of $\alpha$, it can be
written as\cite{Liu07}
\begin{equation}
E(n,\alpha)=E(n,0)+\alpha^{2}[(2^{\frac{2}{3}}-1)<E_{f}^{0}>(u^{\frac{2}{3}}-u)+S_{0}u],
\end{equation}
where $S_{0}=30MeV$ is the bulk symmetry energy parameter. Given the
energy density $\varepsilon(n,\alpha)=n_{0}uE(n,\alpha)$, the
pressure can be expressed as
\begin{equation}
p(n,\alpha)=p(n,0)+n_{0}\alpha^{2}[(2^{\frac{2}{3}}-1)<E_{f}^{0}>(\frac{2}{3}u^{\frac{5}{3}}-u^{2})+S_{0}u^{2}].
\end{equation}

Neglecting the electro-magnetic interaction between the electron and
the nucleon, the energy density and pressure of the neutron star
matter can be expressed as
\begin{equation} \label{et}
\varepsilon_{tot}=\varepsilon_{e}+\varepsilon(n,\alpha),
\end{equation}
and
\begin{equation} \label{pt}
p_{tot}=p_{e}+p(n,\alpha),
\end{equation}
respectively, where $\varepsilon_{e}$ represents the contribution of
the electron to the energy density and\ $p_{e}$\ to the pressure.
\begin{equation}
\varepsilon_{e}(k_{f})=\varepsilon_{0}\int_{0}^{k_{f,e}/m_{e}c}(u^{2}+1)^{\frac{1}{2}}u^{2}du,
\end{equation}
and
\begin{equation}
p_{e}(k_{f})=n\frac{d\varepsilon_{e}}{dn}-\varepsilon_{e}=
\varepsilon_{0}\int_{0}^{k_{f,e}/m_{e}c}(u^{2}+1)^{-\frac{1}{2}}u^{4}du
\end{equation}
with\ $\varepsilon_{0}{=}\frac{m_{e}^{4}c^{5}}{\pi^{2}\hbar^{3}}$.

$\alpha$, n, $k_{f,p}$ and $k_{f,e}$ are all functions of $k_{f,n}$,
so $\varepsilon_{tot}$ and $p_{tot}$ are both functions of
$k_{f,n}$. Although the analytical expression between\
$\varepsilon_{tot}$\ and\ $p_{tot}$\ can not be deduced, the
numerical form of the equation of state can be obtained.

The fourth-order Runge-Kutta algorithm is employed to solve the TOV
equation for the structure of the neutron star.
\begin{equation}
\label{eq:tov}
\frac{dp(r)}{dr}=-\frac{GM(r)\varepsilon(r)}{c^{2}r^{2}}[1+\frac{p(r)}{\varepsilon(r)}][1+\frac{4\pi
r^{3}p(r)}{M(r)c^{2}}][1-\frac{GM(r)}{c^{2}r}],
\end{equation}
and
\begin{equation}
\frac{dM(r)}{dr}=\frac{4\pi r^{2}\varepsilon(r)}{c^{2}},
\end{equation}
where $G$ is the Newtonian gravitation constant, $\varepsilon(r) $
is the energy density at distance $r$ and $p(r)$ the pressure, and
$M(r)$ is the mass of the matter included in the sphere with the
radius $r$. To analyze the dependence of the mass and radius versus
the central pressure, a sequence of central pressures are taken into
the TOV equation. The energy density corresponding to any pressure
is calculated with the linear interpolation method, and the computer
programmes are written with Fortran.

\section{\label{causality}\normalsize amendment of the equation of state}
The square of the rate of  the speed of sound in the nuclear matter
over that of light satisfies \cite{Si03}
\begin{equation}
(\frac{c_{s}}{c})^{2}=\frac{B}{\rho
c^{2}}=\frac{dp}{d\varepsilon}=\frac{dp/du}{d\varepsilon/du}.
\end{equation}

By analysis from Eqs.~(\ref{eq:energypern}) and (\ref{eq:prenm}),
the speed of sound in the nuclear matter will exceed that of light
at high nucleon density, i.e., the equation of state will violate
the relativistic causality in this case. To make it obey the
causality at the high nucleon density, the nuclear potential energy
formula is revised as\cite{Si03}
\begin{equation}
V(n,0)=\frac{A}{2}u+\frac{B}{\sigma+1}\frac{u^{\sigma}}{1+cu^{\sigma-1}},
\end{equation}
where $c$ is dimensionless and when it is set to zero, this formula
returns to the unrevised case. Thereupon, the mean energy per
nucleon for the symmetric matter turns into
\begin{equation}
\frac{\varepsilon(n)}{n}=m_{N}c^{2}+<E_{f}^{0}>u^{\frac{2}{3}}+
\frac{A}{2}u+\frac{B}{\sigma+1}\frac{u^{\sigma}}{1+cu^{\sigma-1}},
\end{equation}
and the pressure changes as
\begin{equation}
p(n)=n_{0}(\frac{2}{3}<E_{f}^{0}>u^{\frac{5}{3}}+
\frac{A}{2}u^{2}+\frac{B}{\sigma+1}\frac{\sigma
u^{\sigma+1}+cu^{2\sigma}}{(1+cu^{\sigma-1})^{2}}).
\end{equation}

The parameters A, B, and $\sigma$ are fixed in the same way as in
Section ~\ref{eos}, but another parameter $c$ has to be determined
firstly. From our calculations, $c$ is suitable between 0.1 and 0.3.
We set $c{=}0.15$ and obtain $\sigma{=}2.246$, $A=-119.468MeV$ and
$B{=}-80.562 MeV$ for the nuclear compressibility $k_0=360MeV$.

Fig.~1 shows the dependence of the energy density versus the nucleon
density of the neutron star matter before and after amendment. At
the low nucleon density, the curves fit together, while at the high
nucleon density, they deviate from each other obviously. As can be
seen in Fig.~1, the equation of state becomes softer after
amendment. As is shown in Fig.~2, the curve of the equation of state
after amendment only approaches the relativistic limit
$p{=}\frac{1}{3}\varepsilon$ and stays distinctly under the causal
limit $p{=}\varepsilon $, which indicates the equation of state
after amendment obeys the causality at the high nucleon density.
\begin{center}
\scalebox {0.85} {\includegraphics{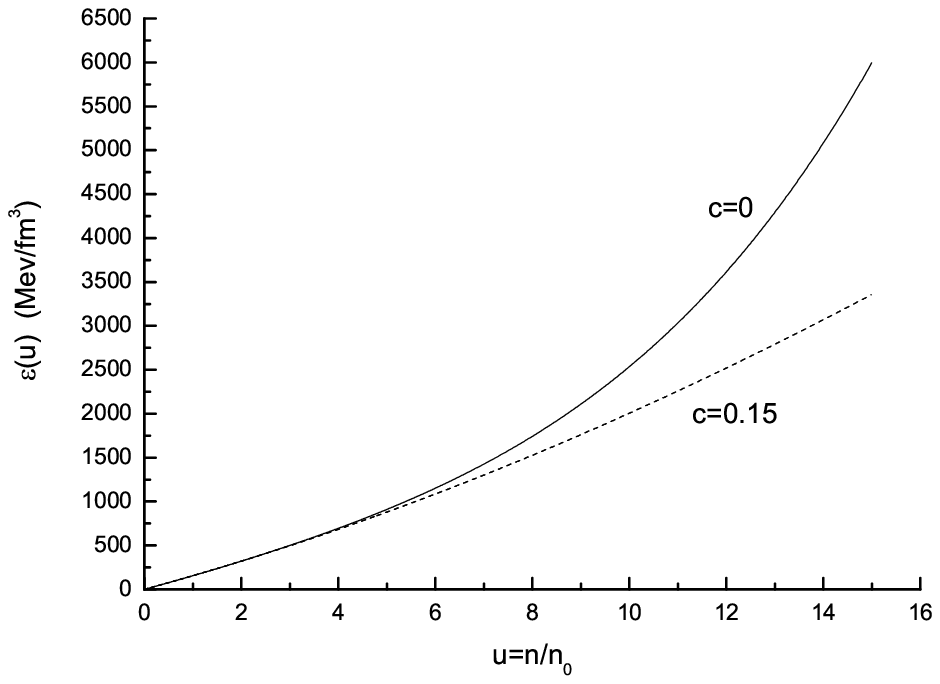}}
\end{center}
\begin{center}
Fig.~1 \ \  The energy density of the $\beta$ stable nuclear matter
versus the nucleon density before and after amendment.
\end{center}
\begin{center}
\scalebox{0.85} {\includegraphics{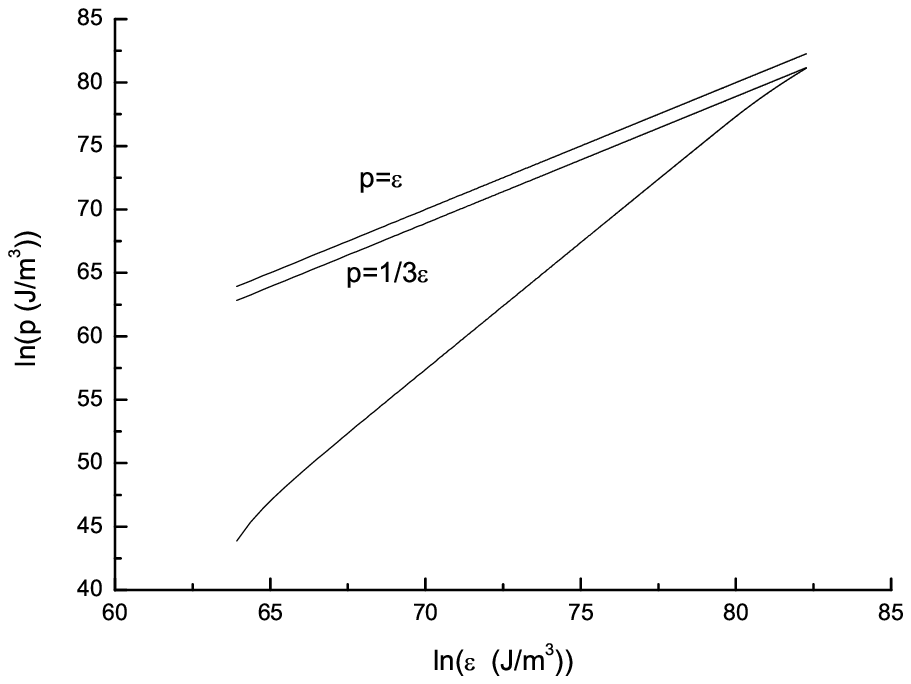}}
\end{center}
\begin{center}
Fig.~2 \ \  The equation of state for the $\beta$ stable nuclear
matter after amendment
\end{center}

\section{\label{result}\normalsize the properties of neutron stars}
By solving the TOV equation, the masses and radii of neutron stars
under different central densities are obtained. In Fig. ~3, the
dashed curve represents the dependence of the radius versus the
central density of neutron stars when $c=0.15$ and the solid one
represents that when $c=0$. Other than minimal deviation at the high
nucleon density, the two curves fit together well basically, so the
revised equation of state has little influence on the radius of the
neutron star. However, the amendment of the equation of state does
make great difference to the relation between the mass and the
central density of neutron stars, as can be seen in Fig.~4, where
the dashed curve is the case of $c=0.15$, and the solid one is the
case of $c=0$. The maximum point of the solid curve corresponds to a
maximum mass $M_{max}=1.932M_{sun}$ with a radius $R=9.370km$, where
$M_{sun} $ is the solar mass, while the one of the dashed curve
corresponds to $M_{max}{=}1.440M_{sun}$ and $R{=}9.765km$.
\begin{center}
\scalebox{0.85} {\includegraphics{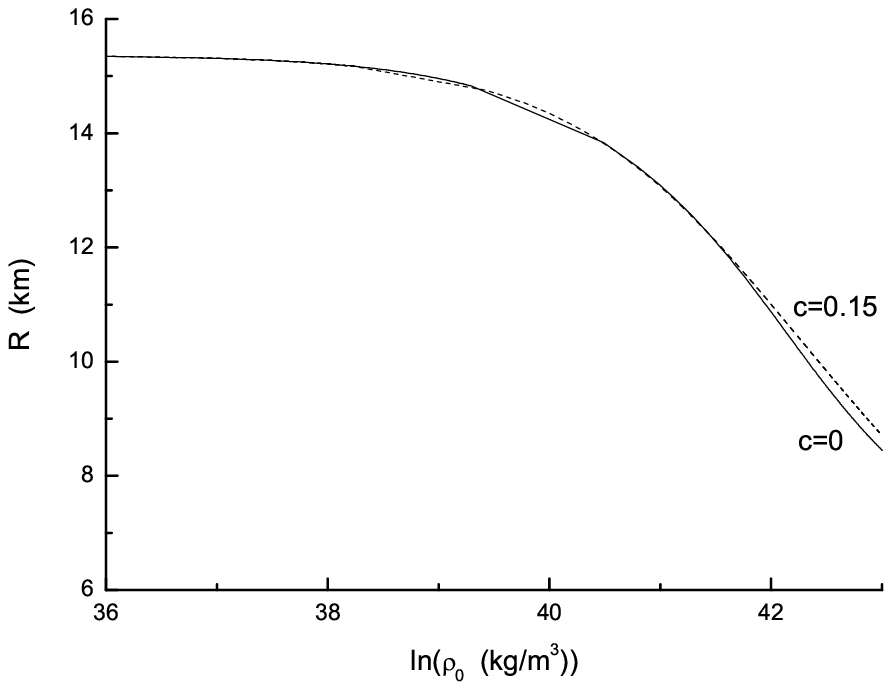}}
\end{center}
\begin{center}
Fig.~3 \ \ The radius versus the central density of neutron stars
  before and after amendment.
\end{center}
\begin{center}
\scalebox{0.85} {\includegraphics{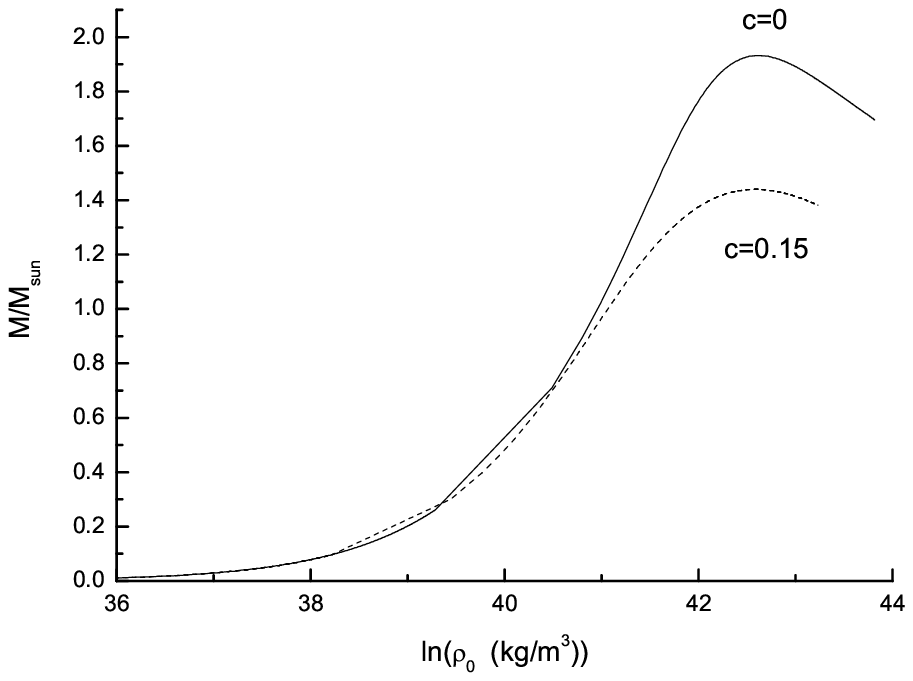}}
\end{center}
\begin{center}
Fig.~4  \ \ The mass versus the central density of neutron stars
  before and after amendment.
\end{center}

\section{\label{cos} \normalsize the influence of the cosmological constant
upon the properties of the neutron star}

If the cosmological constant $\Lambda$ is positive, it has a upper
limit of $4\times10^{-35}s^{-2}$, and if negative, it has a lower
limit of $-2\times10^{-35}s^{-2}$\cite{Oh94}. Considering the
possible influence of $\Lambda$ on the structure of the neutron
star, the factor $[1+\frac{4\pi r^{3}p(r)}{M(r)c^{2}}]$ in the TOV
equation in Eq.~(\ref{eq:tov}) should be changed into $[1+\frac{4\pi
r^{3}p(r)}{M(r)c^{2}}-\frac{\Lambda r^{3}}{2GM(r)}]$.

We take a series of values of $\Lambda$ between the upper and lower
limits into the revised TOV equation and solve it using the equation
of state before the causality amendment. The maximum masses in the
unit of the solar mass and the corresponding radii in the unit of
the kilometer under different values of $\Lambda$ remain unchanged
until the third place behind the decimal point. It implies the
cosmological constant $\Lambda$ has only minimal or even negligible
influence on the properties of the neutron star.

\section{\label{sum}\normalsize summary}
The equation of state for the neutron star matter is calculated
numerically, with the $\beta$ equilibrium conditions considered and
the contribution of the electron included, and then the TOV equation
for the structure of the neutron star is solved. It shows the mass
and radius of the neutron star are functions of the central density.
Considering the equation of state of the nuclear matter must obey
the causality, a new factor $c$ is added in the nuclear potential
energy formula. Therefore, with a new equation of state of the
neutron star matter when $c=0.15$, a maximum mass and the
corresponding radius of the neutron star are obtained. Finally, the
contribution of the cosmological constant to the structure of the
neutron star is discussed, and we find the cosmological constant has
minimal or negligible influence on the properties of the neutron
star.

In the course of working on the graduation thesis on this topic, the
student can learn the knowledge about the degenerate Fermi gas in
quantum mechanics, know about the related contents of the special
and general relativity and has an initial recognition of the
structure of the neutron star. Besides, Chang-Geng Liu has learned
to skillfully write computer programmes with Fortran to solve
physical problems, utilize Origin software to draw figures and edit
scientific articles with Latex. In addition, he has made great
progress in writing and translating scientific articles.


\begin{thebibliography}{50}
\bibitem{Liu07}This formula is the revised version of Eq.~(87) in Ref.~\cite{Si03}.
\end{thebibliography}
\end{document}